\documentclass[a4paper]{jpconf}
\usepackage{amssymb}
\usepackage{amsthm}
\usepackage{latexsym}
\newcommand{\be}{\begin{eqnarray}}
\newcommand{\ee}{\end{eqnarray}}
\newcommand{\bdm}{\begin{displaymath}}
\newcommand{\edm}{\end{displaymath}}

\begin{document}
\title{\textbf{Modified Bargmann-Wigner Formalism (Bosons of Spin 1 and 2)}}


\author{\textbf{Valeri V. Dvoeglazov}}
\address{Universidad de Zacatecas, Apartado Postal 636,
Suc. UAZ\\Zacatecas 98062, Zac., M\'exico\\
E-mail: valeri@planck.reduaz.mx,\,
URL: http://planck.reduaz.mx/\~\,valeri/}


\date{}


\begin{abstract}
On the basis of our recent modifications of the Dirac formalism we 
generalize the Bargmann-Wigner formalism for higher spins to be compatible 
with other formalisms for bosons. Relations with dual electrodynamics, with 
the Ogievetskii-Polubarinov notoph  and the Weinberg 2(2J+1) theory are 
found.  Next, we introduce  the dual analogues of the Riemann tensor and 
derive corresponding dynamical equations in  the Minkowski space. Relations 
with the Marques-Spehler chiral gravity theory are discussed.
\end{abstract}

\section{Introduction}

The equations for higher spins can be derived
from the first principles on using modifications of the Bargmann-Wigner formalism.
The generalizations of the equations in the $(1/2,0)\oplus (0,1/2)$ representation are 
well known. The Tokuoka-SenGupta-Fushchich formalism is based on the equation~\cite{g1,g2,g3,g3a}:
\begin{equation}
[i\gamma_\mu \partial_\mu +m_1 + m_2 \gamma^5] \Psi =0
\end{equation}
If $m_1^2\neq m_2^2$ it was claimed~\cite{g1} that this is simply the change of the representation
of $\gamma$'s. However, the physical consequences are different from those of the Dirac formalism.
Fushchich~\cite{Fush} generalized the formalism
even further in 1970-72, and, in fact, he connected it with the
Gelfand-Tsetlin-Sokolik idea~\cite{gts} of the 2-dimensional
representation of the inversion group. I derived the above parity-violating
equation~\cite{g3a} (and its charge-conjugate) by
the Sakurai-Gersten method from
the first principles. The Barut formalism is based on the equation~\cite{barut,Wilson}:
\begin{equation}
[i\gamma_\mu \partial_\mu +\alpha_2 {\partial_\mu \partial_\mu \over  m} +
\kappa ]
\Psi =0\,.
\end{equation}
It was re-derived from the first principles in~\cite{Dvo,afd-dvo}.
Instead of 4 solutions it has 8 solutions with the correct
relativistic dispersion
$E=\pm \sqrt{{\bf p}^2 +m_i^2}$; and, in fact, it describes
{\it two} mass states
$m_\mu = m_e (1+{3\over 2\alpha})$, provided that the certain physical condition is imposed on the
$\alpha_2$ parameter~\cite{barut}.
One can also generalize the formalism to include the third state,
$\tau$-lepton, see refs.~[7d,10].
Barut also indicated at the possibility of  including $\gamma^5$ term.
For instance, the equation can look something like this:
\begin{equation}
[i\gamma_\mu \partial_\mu +a + b\Box +\gamma^5 (c+d\Box)]\psi =0\,,
\end{equation}
which cannot  be factorized as a product of
two Dirac equations with different masses.

The basic principles of the Weinberg-Tucker-Hammer (WTH) formalism
in the $(J,0)\oplus (0,J)$
representation~\cite{Weinberg,dvo-rmf} are well-known from my previous works.
For spin 1 we have
\begin{equation}
[\gamma_{\alpha\beta} p_\alpha p_\beta +A p_\alpha p_\alpha +Bm^2] \Psi
=0\,, \end{equation} where $p_\mu=-i\partial_\mu$ and
$\gamma_{\alpha\beta}$ are the Barut-Muzinich-Williams covariantly
defined $6\times 6$ matrices~\cite{Bar-Muz}.  The determinant of $[\gamma_{\alpha\beta} p_\alpha
p_\beta +A p_\alpha p_\alpha +Bm^2]$ is of the 12th order in $p_\mu$.
Solutions with $E^2 -{\bf p}^2 = m^2$ can be obtained if and
only if $\frac{B}{A+1} =1\,,\quad \frac{B}{A-1}=1$\,.
The particular cases are:
\begin{itemize}

\item
$A=0, B=1$ $\Leftrightarrow$ we have  the Weinberg's equation for $J=1$
with 3 solutions $E=+\sqrt{{\bf p}^2 +m^2}$, 3 solutions
$E=-\sqrt{{\bf p}^2 +m^2}$, 3 solutions $E=+\sqrt{{\bf p}^2 -m^2}$
and 3 solutions $E=-\sqrt{{\bf p}^2 -m^2}$.

\item
$A=1, B=2$ $\Leftrightarrow$ we have the Tucker-Hammer equation for $J=1$.
The solutions are {\it only}
with $E=\pm\sqrt{{\bf p}^2 +m^2}$.

\end{itemize}

Recently we have shown~\cite{dvo-hpa,dvo-wig}
that one can obtain {\it four} different equations
for antisymmetric tensor fields from the Weinberg $2(2J+1)$
component formalism.  First of all, we note that $\Psi$ is, in fact,
bivector, ${\bf E}_i = -iF_{4i}$, ${\bf B}_i = {1\over 2} \epsilon_{ijk}
F_{jk}$,, or ${\bf E}_i = -{1\over 2} \epsilon_{ijk} \tilde F_{jk}$, ${\bf
B}_i = -i \tilde F_{4i}$, or their combination.  The
four cases are:
\begin{itemize}

\item
$\Psi^{(I)} = \pmatrix{{\bf E} +i{\bf B}\cr
{\bf E} -i{\bf B}\cr}$, $P=-1$, where ${\bf E}_i$ and ${\bf B}_i$ are the
components of the tensor.

\item
$\Psi^{(II)} = \pmatrix{{\bf B} -i{\bf E}\cr
{\bf B} +i{\bf E}\cr}$, $P=+1$, where ${\bf
E}_i$, ${\bf B}_i$ are the components of the tensor.

\item
$\Psi^{(III)} = \Psi^{(I)}$, but (!) ${\bf E}_i$ and ${\bf B}_i$
are the
corresponding vector and axial-vector  components of the
{\it dual} tensor $\tilde F_{\mu\nu}$.

\item
$\Psi^{(IV)} = \Psi^{(II)}$, where ${\bf E}_i$ and ${\bf B}_i$
are the components of the {\it dual} tensor $\tilde F_{\mu\nu}$.

\end{itemize}
The mappings of the WTH equations are:
\begin{eqnarray}
&&\partial_\alpha\partial_\mu F_{\mu\beta}^{(I)}
-\partial_\beta\partial_\mu F_{\mu\alpha}^{(I)}
+ {A-1\over 2} \partial_\mu \partial_\mu F_{\alpha\beta}^{(I)}
-{B\over 2} m^2 F_{\alpha\beta}^{(I)} = 0\,,\label{wth1}\\
&&\partial_\alpha\partial_\mu F_{\mu\beta}^{(II)}
-\partial_\beta\partial_\mu F_{\mu\alpha}^{(II)}
- {A+1\over 2} \partial_\mu \partial_\mu F_{\alpha\beta}^{(II)}
+{B\over 2} m^2 F_{\alpha\beta}^{(II)} = 0\,,\\
&&\partial_\alpha\partial_\mu \tilde F_{\mu\beta}^{(III)}
-\partial_\beta\partial_\mu \tilde F_{\mu\alpha}^{(III)}
- {A+1\over 2} \partial_\mu \partial_\mu \tilde F_{\alpha\beta}^{(III)}
+{B\over 2} m^2 \tilde F_{\alpha\beta}^{(III)} = 0\,,\\
&&\partial_\alpha\partial_\mu \tilde F_{\mu\beta}^{(IV)}
-\partial_\beta\partial_\mu \tilde F_{\mu\alpha}^{(IV)}
+ {A-1\over 2} \partial_\mu \partial_\mu \tilde F_{\alpha\beta}^{(IV)}
-{B\over 2} m^2 \tilde F_{\alpha\beta}^{(IV)} = 0\,.
\end{eqnarray}
In the Tucker-Hammer case ($A=1, B=2$) we can recover the Proca theory
from (\ref{wth1}):
\begin{equation}
\partial_\alpha \partial_\mu F_{\mu\beta}
-\partial_\beta \partial_\mu F_{\mu\alpha} = m^2 F_{\alpha\beta}
\label{proca1}\,.
\end{equation}

Now we are interested in {\it parity-violating}
equations for antisymmetric tensor fields. We also study
the most general mapping of the Weinberg-Tucker-Hammer formulation
to the antisymetric tensor field formulation.
Instead of $\Psi^{(I-IV)}$ we shall try to use now
\begin{equation}
\Psi^{(A)} = \pmatrix{{\bf E} +i{\bf B}\cr
{\bf B} +i{\bf E}\cr} = {1+\gamma^5\over 2} \Psi^{(I)}+
{1-\gamma^5 \over 2} \Psi^{(II)}\,.
\end{equation}
As a result, the equation for the AST fields is
\begin{equation}
\partial_\alpha \partial_\mu F_{\mu\beta}
-\partial_\beta \partial_\mu F_{\mu\alpha}
={1\over 2} (\partial_\mu \partial_\mu) F_{\alpha\beta} +
[-{A\over 2} (\partial_\mu \partial_\mu) + {B\over 2} m^2] \tilde
F_{\alpha\beta}\,.\label{pv1}
\end{equation}
The different choice is
\begin{equation} \Psi^{(B)} = \pmatrix{{\bf E} +i{\bf
B}\cr -{\bf B} -i{\bf E}\cr} = {1+\gamma^5\over 2} \Psi^{(I)}- {1-\gamma^5
\over 2} \Psi^{(II)}\,.
\end{equation}
Thus, one has
\begin{equation}
\partial_\alpha \partial_\mu F_{\mu\beta}
-\partial_\beta \partial_\mu F_{\mu\alpha}
={1\over 2} (\partial_\mu \partial_\mu) F_{\alpha\beta} +
[{A\over 2} (\partial_\mu \partial_\mu)- {B\over 2} m^2] \tilde
F_{\alpha\beta}\,.\label{pv2}
\end{equation}
Of course, one can also
use the dual tensor and obtain analogous equations:
\begin{eqnarray}
&&\partial_\alpha \partial_\mu \tilde F_{\mu\beta}
-\partial_\beta \partial_\mu \tilde F_{\mu\alpha}
={1\over 2} (\partial_\mu \partial_\mu) \tilde F_{\alpha\beta} +
[-{A\over 2} (\partial_\mu \partial_\mu) + {B\over 2} m^2]
F_{\alpha\beta}\,,\\
&&\partial_\alpha \partial_\mu \tilde F_{\mu\beta}
-\partial_\beta \partial_\mu \tilde F_{\mu\alpha}
={1\over 2} (\partial_\mu \partial_\mu) \tilde F_{\alpha\beta} +
[{A\over 2} (\partial_\mu \partial_\mu) - {B\over 2} m^2]
F_{\alpha\beta}\,.  \end{eqnarray}
They are connected
with (\ref{pv1},\ref{pv2}) by the dual transformations.

The states corresponding to the new functions $\Psi^{(A)}$,
$\Psi^{(B)}$ etc are {\it not} the parity eigenstates. So, it is not
surprising that we have $F_{\alpha\beta}$ and
its dual $\tilde F_{\alpha\beta}$ in
the same equations. In total we have already eight equations.

One can also consider the most general case
\begin{equation}
\Psi^{(W)} =\pmatrix{aF_{4i} +b \tilde F_{4i} +c \epsilon_{ijk} F_{jk}
+d \epsilon_{ijk} \tilde F_{jk}\cr
eF_{4i} +f \tilde F_{4i} +g \epsilon_{ijk} F_{jk}
+h \epsilon_{ijk} \tilde F_{jk}\cr}\,.
\end{equation}
So, we have dynamical equations for $F_{\alpha\beta}$
and $\tilde F_{\alpha\beta}$ with additional parameters $a,b,c,d,\ldots
\in\, {\bf C}$. We have a lot of antisymmetric tensor fields here.

The Bargmann-Wigner formalism
for constructing  of high-spin particles has been given
in~\cite{bw-hs,Lurie}. However, they claimed
explicitly that they constructed $(2J+1)$ states (the
Weinberg-Tucker-Hammer theory has  essentially $2(2J+1)$
components).  The standard Bargmann-Wigner formalism for $J=1$ is based on
the following set 
\begin{eqnarray} \left [ i\gamma_\mu \partial_\mu +m \right
]_{\alpha\beta} \Psi_{\beta\gamma} &=& 0\,,\label{bw1}\\ \left [
i\gamma_\mu \partial_\mu +m \right ]_{\gamma\beta} \Psi_{\alpha\beta} &=&
0\,, \label{bw2} \end{eqnarray} 
If one has
\begin{equation} \Psi_{\left \{ \alpha\beta
\right \} } = (\gamma_\mu R)_{\alpha\beta} A_\mu +
(\sigma_{\mu\nu} R)_{\alpha\beta} F_{\mu\nu}\,,
\end{equation} with
\begin{equation} 
R = e^{i\varphi}
\pmatrix{\Theta&0\cr 0&-\Theta\cr}\,\quad
\Theta=\pmatrix{0&-1\cr
1&0\cr}
\end{equation} in the spinorial
representation of $\gamma$-matrices we obtain
the Duffin-Proca-Kemmer equations:
\begin{eqnarray}
&&\partial_\alpha F_{\alpha\mu} = {m\over 2} A_\mu\,,\\
&& 2m F_{\mu\nu} = \partial_\mu A_\nu - \partial_\nu A_\mu\,.
\end{eqnarray}
After the corresponding
re-normalization $A_\mu \rightarrow 2m A_\mu$, we obtain the standard
textbook set:
\begin{eqnarray} &&\partial_\alpha
F_{\alpha\mu} = m^2 A_\mu\,,\\ && F_{\mu\nu} = \partial_\mu A_\nu -
\partial_\nu A_\mu\,.  \end{eqnarray} 
It gives the equation (\ref{proca1})
for the antisymmetric tensor field. How can one obtain other equations
following the Weinberg-Tucker-Hammer approach?
The third equation can be obtained in a  simple way: use, instead of
$(\sigma_{\mu\nu} R) F_{\mu\nu}$, another symmetric matrix $(\gamma^5
\sigma_{\mu\nu} R)  F_{\mu\nu}$, see~\cite{dv-ps}.
And what about the second and the fourth equations?  I suggest:

\begin{itemize}

\item
to use, see above and~\cite{g1}:
\begin{equation}
[i\gamma_\mu \partial_\mu +m ] \Psi =0 \Rightarrow
[i\gamma_\mu \partial_\mu +m_1 +m_2\gamma_5 ] \Psi =0\,;
\end{equation}

\item
to use the Barut extension:
\begin{equation}
[i\gamma_\mu \partial_\mu +m ] \Psi =0 \Rightarrow
[i\gamma_\mu \partial_\mu +a{\partial_\mu \partial_\mu\over m} +\kappa ] \Psi
=0\,.
\end{equation}
In such a way we can enlarge the set of possible states.

\end{itemize}

\section{Modified Bargmann-Wigner Formalism}

We begin with
\begin{eqnarray}
\left [ i\gamma_\mu \partial_\mu + a -b \Box + \gamma_5 (c- d\Box )
\right ]_{\alpha\beta} \Psi_{\beta\gamma} &=&0\,,\\
\left [ i\gamma_\mu
\partial_\mu + a -b \Box - \gamma_5 (c- d\Box ) \right ]_{\alpha\beta}
\Psi_{\gamma\beta} &=&0\,,
\end{eqnarray}
$\Box$ is the d'Alembertian.
Thus, we obtain the Proca-like equations:
\begin{eqnarray} &&\partial_\nu A_\lambda - \partial_\lambda A_\nu - 2(a
+b \partial_\mu \partial_\mu ) F_{\nu \lambda} =0\,,\\ &&\partial_\mu
F_{\mu \lambda} = {1\over 2} (a +b \partial_\mu \partial_\mu) A_\lambda +
{1\over 2} (c+ d \partial_\mu \partial_\mu) \tilde A_\lambda\,,
\end{eqnarray}
$\tilde A_\lambda$ is the axial-vector potential (analogous to that
used in the Duffin-Kemmer set for $J=0$). Additional constraints are:
\begin{eqnarray}
&&i\partial_\lambda A_\lambda + ( c+d\partial_\mu \partial_\mu) \tilde \phi
=0\,,\\
&&\epsilon_{\mu\lambda\kappa \tau} \partial_\mu F_{\lambda\kappa } =0\,,
( c+ d \partial_\mu \partial_\mu ) \phi =0\,.
\end{eqnarray}

The spin-0 Duffin-Kemmer equations are:
\begin{eqnarray}
&&(a+b \partial_\mu \partial_\mu) \phi = 0\,, 
i\partial_\mu \tilde A_\mu  - (a+b\partial_\mu \partial_\mu) \tilde
\phi =0\,,\\
&&(a+b\, \partial_\mu \partial_\mu) \tilde A_\nu + (c+d\,\partial_\mu
\partial_\mu) A_\nu + i (\partial_\nu \tilde \phi) =0\,.
\end{eqnarray}
The additional constraints are:
\begin{equation}
\partial_\mu \phi =0\,,
\partial_\nu \tilde A_\lambda - \partial_\lambda \tilde
A_\nu +2 (c+d\partial_\mu \partial_\mu ) F_{\nu \lambda} = 0\,.
\end{equation}
In such a way the spin states are {\it mixed} through the 4-vector potentials.
After elimination of the 4-vector potentials we obtain
the equation for the AST field of the second rank:
\begin{equation}
\left [ \partial_\mu \partial_\nu F_{\nu\lambda} - \partial_\lambda
\partial_\nu F_{\nu\mu}\right ]   + \left [ (c^2 - a^2) - 2(ab-cd)
\partial_\mu\partial_\mu  + (d^2 -b^2)
(\partial_\mu\partial_\mu)^2 \right ] F_{\mu\lambda} = 0\,,
\end{equation}
which should be compared with our
previous equations which follow from the Weinberg-like formulation.
Just put:
\begin{eqnarray}
c^2 - a^2 \Rightarrow {-Bm^2 \over 2}\,,&\qquad& c^2 - a^2 \Rightarrow
+{Bm^2 \over 2}\,,\\
-2(ab-cd) \Rightarrow {A-1\over 2}\,,&\qquad&
+2(ab-cd) \Rightarrow {A+1\over 2}\,,\\
b=\pm d\,.&\qquad&
\end{eqnarray}
Of course, these sets of algebraic equations have solutions in terms $A$
and $B$. We found them and restored the equations, see above.

The parity violation and the spin mixing are {\it intrinsic} possibilities
of the Proca-like theories. One can go in a different way:
instead of modifying the equations, consider the spin basis rotations.
In the helicity basis we have (see also~\cite{Ber,Grei}, where it was
claimed explicitly that helicity states cannot be parity eigenstates):
\begin{eqnarray}
&&\epsilon _{\mu }({\bf p},\lambda =+1)={\frac{1}{\sqrt{2}}}{\frac{e^{i\phi }}{
p}}\pmatrix{ 0, {p_x p_z -ip_y p\over \sqrt{p_x^2 +p_y^2}}, {p_y p_z +ip_x
p\over \sqrt{p_x^2 +p_y^2}}, -\sqrt{p_x^2 +p_y^2}}\,, \\
&&\epsilon _{\mu }({\bf p},\lambda =-1)={\frac{1}{\sqrt{2}}}{\frac{e^{-i\phi }
}{p}}\pmatrix{ 0, {-p_x p_z -ip_y p\over \sqrt{p_x^2 +p_y^2}}, {-p_y p_z
+ip_x p\over \sqrt{p_x^2 +p_y^2}}, +\sqrt{p_x^2 +p_y^2}}\,, \\
&&\epsilon _{\mu }({\bf p},\lambda =0)={\frac{1}{m}}\pmatrix{ p, -{E \over p}
p_x, -{E \over p} p_y, -{E \over p} p_z }\,, 
\epsilon _{\mu }({\bf p},\lambda =0_{t})={\frac{1}{m}}\pmatrix{E , -p_x,
-p_y, -p_z }.
\end{eqnarray}
and
\begin{eqnarray}
&&{\bf E}({\bf p},\lambda =+1)=-{\frac{iEp_{z}}{\sqrt{2}pp_{l}}}{\bf p}-{\frac{
E}{\sqrt{2}p_{l}}}\tilde{{\bf p}},\quad {\bf B}({\bf p},\lambda =+1)=-{\frac{
p_{z}}{\sqrt{2}p_{l}}}{\bf p}+{\frac{ip}{\sqrt{2}p_{l}}}\tilde{{\bf p}}, \\
&&{\bf E}({\bf p},\lambda =-1)=+{\frac{iEp_{z}}{\sqrt{2}pp_{r}}}{\bf p}-{\frac{
E}{\sqrt{2}p_{r}}}\tilde{{\bf p}}^{\ast },\quad {\bf B}({\bf p},\lambda
=-1)=-{\frac{p_{z}}{\sqrt{2}p_{r}}}{\bf p}-{\frac{ip}{\sqrt{2}p_{r}}}\tilde{
{\bf p}}^{\ast }, \\
&&{\bf E}({\bf p},\lambda =0)={\frac{im}{p}}{\bf p},\quad {\bf B}({\bf p}
,\lambda =0)=0,
\end{eqnarray}
with $\tilde {\bf p}=column (p_y\,, -p_x\,, -ip )$.

In fact, there  are several modifications of the BW formalism. One can came to the following set:
\begin{eqnarray}
\left [ i\gamma_\mu \partial_\mu + \epsilon_1 m_1 +\epsilon_2 m_2 \gamma_5
\right ]_{\alpha\beta} \Psi_{\beta\gamma} &=&0\,,\\
\left [ i\gamma_\mu
\partial_\mu + \epsilon_3 m_1 +\epsilon_4 m_2 \gamma_5 \right ]_{\alpha\beta}
\Psi_{\gamma\beta} &=&0\,,
\end{eqnarray}
where $\epsilon_i$ are the sign operators. So, at first sight, we have 16
possible combinations for the AST fields. We first come to
\begin{eqnarray}
&&\left [ i\gamma_\mu \partial_\mu + m_1 A_1 + m_2 A_2\gamma_5
\right ]_{\alpha\beta} \left \{ (\gamma_\lambda R)_{\beta\gamma} A_\lambda
+ (\sigma_{\lambda\kappa } R)_{\beta\gamma} F_{\lambda\kappa }\right
\}+\nonumber\\ &+&\left [ m_1 B_1 +m_2 B_2 \gamma_5 \right ] \left \{
R_{\beta\gamma}\varphi + (\gamma_5 R)_{\beta\gamma} \tilde \phi +(\gamma_5
\gamma_\lambda R)_{\beta\gamma} \tilde A_\lambda\right \}=0\,,\\
&&\left [
i\gamma_\mu \partial_\mu + m_1 A_1 + m_2 A_2\gamma_5 \right
]_{\gamma\beta} \left \{ (\gamma_\lambda R)_{\alpha\beta} A_\lambda +
(\sigma_{\lambda\kappa } R)_{\alpha\beta} F_{\lambda\kappa }\right \}-\nonumber\\
&-&\left [ m_1 B_1 +m_2 B_2 \gamma_5 \right ] \left \{
R_{\alpha\beta}\varphi +(\gamma_5 R)_{\alpha\beta} \tilde \phi +(\gamma_5
\gamma_\lambda R)_{\alpha\beta} \tilde A_\lambda\right \}=0\,,
\end{eqnarray}
where $A_1 = {\epsilon_1 +\epsilon_3 \over 2}$,
$A_2 = {\epsilon_2 +\epsilon_4 \over 2}$,
$B_1 = {\epsilon_1 -\epsilon_3 \over 2}$,
and
$B_2 = {\epsilon_2 -\epsilon_4 \over 2}$.
Thus for spin 1 we have
\begin{eqnarray} &&\partial_\mu A_\lambda - \partial_\lambda A_\mu + 2m_1 A_1 F_{\mu \lambda}
+im_2 A_2 \epsilon_{\alpha\beta\mu\lambda} F_{\alpha\beta} =0\,,\\
&&\partial_\lambda
F_{\kappa \lambda} - {m_1\over 2} A_1 A_\kappa -{m_2\over 2} B_2 \tilde
A_\kappa=0\,,
\end{eqnarray}  with constraints
\begin{eqnarray}
&&-i\partial_\mu A_\mu + 2m_1 B_1 \phi +2m_2 B_2 \tilde \phi=0\,,\\
&&i\epsilon_{\mu\nu\kappa\lambda} \partial_\mu F_{\nu\kappa}
-m_2 A_2 A_\lambda -m_1 B_1 \tilde A_\lambda =0\,,\\
&&m_1 B_1 \tilde \phi +m_2 B_2 \phi =0\,.
\end{eqnarray}
If we remove $A_\lambda$ and $\tilde A_\lambda$ from this set,
we come to the final results for the AST field.
Actually, we have twelve equations, see~\cite{dvo-wig}. One can
go even further. One can use the Barut equations for the BW input. So, we
can get $16\times 16$ combinations (depending on the eigenvalues of the
corresponding sign operators), and we have different eigenvalues of masses 
due to $\partial_\mu^2 = \kappa m^2$.

Why do I think that the shown arbitrarieness
of equations for the AST fields is related to 1) spin basis rotations; 2)
the choice of normalization?  In the common-used basis the three
4-potentials have parity eigenvalues $-1$ and one time-like (or spin-0
state), $+1$; the fields ${\bf E}$ and ${\bf B}$ have also definite parity
properties in this basis.  If we transfer to other  basis, e.g., to the
helicity basis we can see that the 4-vector potentials and
the corresponding fields are superpositions of the vector and the
axial-vector.  Of course, they can be expanded in the fields in the
``old" basis. 

So, we conclude:
the addition of the Klein-Gordon equation to the $(J,0)\oplus (0,J)$
equations may change physical content even on the free level.
In the $(1/2,0)\oplus (0,1/2)$ representation it is possible to introduce
the {\it parity-violating} frameworks.
We found the mappings between the Weinberg-Tucker-Hammer formalism for
$J=1$ and the AST fields of the 2nd rank of at least eight types. Four
of them include both $F_{\mu\nu}$ and $\tilde F_{\mu\nu}$, which tells us
that the parity violation may occur during  the study of the corresponding
dynamics.
If we want to take into account the $J=1$ solutions with
different parity properties, the Bargmann-Wigner (BW), the Proca and the
Duffin-Kemmer-Petiau (DKP)
formalisms are to be generalized.
We considered the most general
case, introducing eight scalar parameters. In order to have covariant
equations for the AST fields, one should impose constraints on the
corresponding parameters.
It is possible to get solutions with mass
splitting.
We found the 4-potentials and fields in the helicity
basis. They have different parity properties comparing with the standard
(``parity") basis (cf.~\cite{Ber,Grei}).
The discussion induced us to generalize the BW,
the Proca and the Duffin-Kemmer-Petiau formalisms.  Higher-spin equations
may actually describe various spin, mass, helicity and parity states.  The
states of different parity, helicity, and mass may be present in the same
equation.
On the basis of generalizations of the BW formalism, finally,  we obtained
{\it twelve} equations for the AST fields.
A hypothesis was
presented that the obtained results are related to the spin basis
rotations and to the choice of normalization.

\section{Standard Formalism (Spin 2)}

The general scheme for derivation of higher-spin equations
was given in~\cite{bw-hs}. A field of rest mass $m$ and spin $j \geq {1\over
2}$ is represented by a completely symmetric multispinor of rank $2j$.
The particular cases $j=1$ and $j={3\over 2}$ were given in the
textbooks, e.~g., ref.~\cite{Lurie}. The spin-2 case can also be of some
interest because it is generally believed that the essential features of
the gravitational field are  obtained from transverse components of the
$(2,0)\oplus (0,2)$  representation of the Lorentz group. Nevertheless,
questions of the redandant components of the higher-spin relativistic
equations are not yet understood in detail~\cite{Kirch}.

In this section we use the commonly-accepted procedure
for the derivation  of higher-spin equations.
We begin with the equations for the 4-rank symmetric spinor:
\begin{eqnarray}
&&\left [ i\gamma^\mu \partial_\mu - m \right ]_{\alpha\alpha^\prime}
\Psi_{\alpha^\prime \beta\gamma\delta} = 0\, ,
\left [ i\gamma^\mu \partial_\mu - m \right ]_{\beta\beta^\prime}
\Psi_{\alpha\beta^\prime \gamma\delta} = 0\, ,\\
&&\left [ i\gamma^\mu \partial_\mu - m \right ]_{\gamma\gamma^\prime}
\Psi_{\alpha\beta\gamma^\prime \delta} = 0\, ,
\left [ i\gamma^\mu \partial_\mu - m \right ]_{\delta\delta^\prime}
\Psi_{\alpha\beta\gamma\delta^\prime} = 0\, .
\end{eqnarray} 
The massless limit (if one needs) should be taken in the end of all
calculations.

We proceed expanding the field function in the set of symmetric matrices
(as in the spin-1 case, cf.~ref.~[4a]). In the beginning let us use the
first two indices:
\begin{equation} \Psi_{\{\alpha\beta\}\gamma\delta} =
(\gamma_\mu R)_{\alpha\beta} \Psi^\mu_{\gamma\delta}
+(\sigma_{\mu\nu} R)_{\alpha\beta} \Psi^{\mu\nu}_{\gamma\delta}\, .
\end{equation}
We would like to write
the corresponding equations for functions $\Psi^\mu_{\gamma\delta}$
and $\Psi^{\mu\nu}_{\gamma\delta}$ in the form:
\begin{equation}
{2\over m} \partial_\mu \Psi^{\mu\nu}_{\gamma\delta} = -
\Psi^\nu_{\gamma\delta}\, , 
\Psi^{\mu\nu}_{\gamma\delta} = {1\over 2m}
\left [ \partial^\mu \Psi^\nu_{\gamma\delta} - \partial^\nu
\Psi^\mu_{\gamma\delta} \right ]\, \label{p2}.
\end{equation} 
Constraints $(1/m) \partial_\mu \Psi^\mu_{\gamma\delta} =0$
and $(1/m) \epsilon^{\mu\nu}_{\quad\alpha\beta}\, \partial_\mu
\Psi^{\alpha\beta}_{\gamma\delta} = 0$ can be regarded as a consequence of
Eqs.  (\ref{p2}).
Next, we present the vector-spinor and tensor-spinor functions as
\begin{eqnarray}
&&\Psi^\mu_{\{\gamma\delta\}} = (\gamma^\kappa R)_{\gamma\delta}
G_{\kappa}^{\quad \mu} +(\sigma^{\kappa\tau} R )_{\gamma\delta}
F_{\kappa\tau}^{\quad \mu} \, ,\\
&&\Psi^{\mu\nu}_{\{\gamma\delta\}} = (\gamma^\kappa R)_{\gamma\delta}
T_{\kappa}^{\quad \mu\nu} +(\sigma^{\kappa\tau} R )_{\gamma\delta}
R_{\kappa\tau}^{\quad \mu\nu} \, ,
\end{eqnarray}
i.~e.,  using the symmetric matrix coefficients in indices $\gamma$ and
$\delta$. Hence, the total function is
\begin{eqnarray}
\lefteqn{\Psi_{\{\alpha\beta\}\{\gamma\delta\}}
= (\gamma_\mu R)_{\alpha\beta} (\gamma^\kappa R)_{\gamma\delta}
G_\kappa^{\quad \mu} + (\gamma_\mu R)_{\alpha\beta} (\sigma^{\kappa\tau}
R)_{\gamma\delta} F_{\kappa\tau}^{\quad \mu} + } \nonumber\\
&+& (\sigma_{\mu\nu} R)_{\alpha\beta} (\gamma^\kappa R)_{\gamma\delta}
T_\kappa^{\quad \mu\nu} + (\sigma_{\mu\nu} R)_{\alpha\beta}
(\sigma^{\kappa\tau} R)_{\gamma\delta} R_{\kappa\tau}^{\quad\mu\nu} \, ;
\end{eqnarray}
and the resulting tensor equations are:
\begin{eqnarray}
&&{2\over m} \partial_\mu T_\kappa^{\quad \mu\nu} =
-G_{\kappa}^{\quad\nu}\, ,
{2\over m} \partial_\mu R_{\kappa\tau}^{\quad \mu\nu} =
-F_{\kappa\tau}^{\quad\nu}\, ,\\
&& T_{\kappa}^{\quad \mu\nu} = {1\over 2m} \left [
\partial^\mu G_{\kappa}^{\quad\nu}
- \partial^\nu G_{\kappa}^{\quad \mu} \right ] \, ,\\
&& R_{\kappa\tau}^{\quad \mu\nu} = {1\over 2m} \left [
\partial^\mu F_{\kappa\tau}^{\quad\nu}
- \partial^\nu F_{\kappa\tau}^{\quad \mu} \right ] \, .
\end{eqnarray}
The constraints are re-written to
\begin{eqnarray}
&&{1\over m} \partial_\mu G_\kappa^{\quad\mu} = 0\, ,\quad
{1\over m} \partial_\mu F_{\kappa\tau}^{\quad\mu} =0\, ,\\
&& {1\over m} \epsilon_{\alpha\beta\nu\mu} \partial^\alpha
T_\kappa^{\quad\beta\nu} = 0\, ,\quad
{1\over m} \epsilon_{\alpha\beta\nu\mu} \partial^\alpha
R_{\kappa\tau}^{\quad\beta\nu} = 0\, .
\end{eqnarray}
However, we need to make symmetrization over these two sets
of indices $\{ \alpha\beta \}$ and $\{\gamma\delta \}$. The total
symmetry can be ensured if one contracts the function $\Psi_{\{\alpha\beta
\} \{\gamma \delta \}}$ with {\it antisymmetric} matrices
$R^{-1}_{\beta\gamma}$, $(R^{-1} \gamma^5 )_{\beta\gamma}$ and
$(R^{-1} \gamma^5 \gamma^\lambda )_{\beta\gamma}$ and equate
all these contractions to zero (similar to the $j=3/2$ case
considered in ref.~\cite[p. 44]{Lurie}. We obtain
additional constraints on the tensor field functions:
\begin{eqnarray}
&& G_\mu^{\quad\mu}=0\, , \quad G_{[\kappa \, \mu ]}  = 0\, , \quad
G^{\kappa\mu} = {1\over 2} g^{\kappa\mu} G_\nu^{\quad\nu}\, ,
\label{b1}\\
&&F_{\kappa\mu}^{\quad\mu} = F_{\mu\kappa}^{\quad\mu} = 0\, , \quad
\epsilon^{\kappa\tau\mu\nu} F_{\kappa\tau,\mu} = 0\, ,\\
&& T^{\mu}_{\quad\mu\kappa} =
T^{\mu}_{\quad\kappa\mu} = 0\, ,\quad
\epsilon^{\kappa\tau\mu\nu} T_{\kappa,\tau\mu} = 0\, ,\\
&& F^{\kappa\tau,\mu} = T^{\mu,\kappa\tau}\, ,\quad
\epsilon^{\kappa\tau\mu\lambda} (F_{\kappa\tau,\mu} +
T_{\kappa,\tau\mu})=0\, ,\\
&& R_{\kappa\nu}^{\quad \mu\nu}
= R_{\nu\kappa}^{\quad  \mu\nu} = R_{\kappa\nu}^{\quad\nu\mu}
= R_{\nu\kappa}^{\quad\nu\mu}
= R_{\mu\nu}^{\quad  \mu\nu} = 0\, , \\
&& \epsilon^{\mu\nu\alpha\beta} (g_{\beta\kappa} R_{\mu\tau,
\nu\alpha} - g_{\beta\tau} R_{\nu\alpha,\mu\kappa} ) = 0\, \quad
\epsilon^{\kappa\tau\mu\nu} R_{\kappa\tau,\mu\nu} = 0\, .\label{f1}
\end{eqnarray} 
Thus, we  encountered with
the known difficulty of the theory for spin-2 particles in
the Minkowski space.
We explicitly showed that all field functions become to be equal to zero.
Such a situation cannot be considered as a satisfactory one (because it
does not give us any physical information) and can be corrected in several
ways.\footnote{The reader can compare our results of this Section with
those of G. Marques and D. Spehler, Mod. Phys. Lett. A13 (1998) 553-569.
I consider their discussion
of the standard formalism in the Sections I and II, as insufficient.}

\section{Generalized Formalism (Spin 2)}

We shall modify the formalism~\cite{dv-ps}. The field function is now presented as
\begin{equation}
\Psi_{\{\alpha\beta\}\gamma\delta} =
\alpha_1 (\gamma_\mu R)_{\alpha\beta} \Psi^\mu_{\gamma\delta} +
\alpha_2 (\sigma_{\mu\nu} R)_{\alpha\beta} \Psi^{\mu\nu}_{\gamma\delta}
+\alpha_3 (\gamma^5 \sigma_{\mu\nu} R)_{\alpha\beta}
\widetilde \Psi^{\mu\nu}_{\gamma\delta}\, ,
\end{equation}
with
\begin{eqnarray}
&&\Psi^\mu_{\{\gamma\delta\}} = \beta_1 (\gamma^\kappa R)_{\gamma\delta}
G_\kappa^{\quad\mu} + \beta_2 (\sigma^{\kappa\tau} R)_{\gamma\delta}
F_{\kappa\tau}^{\quad\mu} +\beta_3 (\gamma^5 \sigma^{\kappa\tau}
R)_{\gamma\delta} \widetilde F_{\kappa\tau}^{\quad\mu} \, ,\\
&&\Psi^{\mu\nu}_{\{\gamma\delta\}} =\beta_4 (\gamma^\kappa
R)_{\gamma\delta} T_\kappa^{\quad\mu\nu} + \beta_5 (\sigma^{\kappa\tau}
R)_{\gamma\delta} R_{\kappa\tau}^{\quad\mu\nu} +\beta_6 (\gamma^5
\sigma^{\kappa\tau} R)_{\gamma\delta}
\widetilde R_{\kappa\tau}^{\quad\mu\nu} \, ,\\
&&\widetilde \Psi^{\mu\nu}_{\{\gamma\delta\}} =\beta_7 (\gamma^\kappa
R)_{\gamma\delta} \widetilde T_\kappa^{\quad\mu\nu} + \beta_8
(\sigma^{\kappa\tau} R)_{\gamma\delta}
\widetilde D_{\kappa\tau}^{\quad\mu\nu}
+\beta_9 (\gamma^5 \sigma^{\kappa\tau} R)_{\gamma\delta}
D_{\kappa\tau}^{\quad\mu\nu} \, .
\end{eqnarray}
Hence, the function $\Psi_{\{\alpha\beta\}\{\gamma\delta\}}$
can be expressed as a sum of nine terms:
\begin{eqnarray}
&&\Psi_{\{\alpha\beta\}\{\gamma\delta\}} =
\alpha_1 \beta_1 (\gamma_\mu R)_{\alpha\beta} (\gamma^\kappa
R)_{\gamma\delta} G_\kappa^{\quad\mu} +\alpha_1 \beta_2
(\gamma_\mu R)_{\alpha\beta} (\sigma^{\kappa\tau} R)_{\gamma\delta}
F_{\kappa\tau}^{\quad\mu} + \nonumber\\
&+&\alpha_1 \beta_3 (\gamma_\mu R)_{\alpha\beta}
(\gamma^5 \sigma^{\kappa\tau} R)_{\gamma\delta} \widetilde
F_{\kappa\tau}^{\quad\mu} +
+ \alpha_2 \beta_4 (\sigma_{\mu\nu}
R)_{\alpha\beta} (\gamma^\kappa R)_{\gamma\delta} T_\kappa^{\quad\mu\nu}
+\nonumber\\
&+&\alpha_2 \beta_5 (\sigma_{\mu\nu} R)_{\alpha\beta} (\sigma^{\kappa\tau}
R)_{\gamma\delta} R_{\kappa\tau}^{\quad \mu\nu}
+ \alpha_2
\beta_6 (\sigma_{\mu\nu} R)_{\alpha\beta} (\gamma^5 \sigma^{\kappa\tau}
R)_{\gamma\delta} \widetilde R_{\kappa\tau}^{\quad\mu\nu} +\nonumber\\
&+&\alpha_3 \beta_7 (\gamma^5 \sigma_{\mu\nu} R)_{\alpha\beta}
(\gamma^\kappa R)_{\gamma\delta} \widetilde
T_\kappa^{\quad\mu\nu}+
\alpha_3 \beta_8 (\gamma^5
\sigma_{\mu\nu} R)_{\alpha\beta} (\sigma^{\kappa\tau} R)_{\gamma\delta}
\widetilde D_{\kappa\tau}^{\quad\mu\nu} +\nonumber\\
&+&\alpha_3 \beta_9
(\gamma^5 \sigma_{\mu\nu} R)_{\alpha\beta} (\gamma^5 \sigma^{\kappa\tau}
R)_{\gamma\delta} D_{\kappa\tau}^{\quad \mu\nu}\, .
\label{ffn1}
\end{eqnarray}
The corresponding dynamical
equations are given by the set
\begin{eqnarray}
&& {2\alpha_2
\beta_4 \over m} \partial_\nu T_\kappa^{\quad\mu\nu} +{i\alpha_3
\beta_7 \over m} \epsilon^{\mu\nu\alpha\beta} \partial_\nu
\widetilde T_{\kappa,\alpha\beta} = \alpha_1 \beta_1
G_\kappa^{\quad\mu}\,; \label{b}\\
&&{2\alpha_2 \beta_5 \over m} \partial_\nu
R_{\kappa\tau}^{\quad\mu\nu} +{i\alpha_2 \beta_6 \over m}
\epsilon_{\alpha\beta\kappa\tau} \partial_\nu \widetilde R^{\alpha\beta,
\mu\nu} +{i\alpha_3 \beta_8 \over m}
\epsilon^{\mu\nu\alpha\beta}\partial_\nu \widetilde
D_{\kappa\tau,\alpha\beta} - \nonumber\\
&-&{\alpha_3 \beta_9 \over 2}
\epsilon^{\mu\nu\alpha\beta} \epsilon_{\lambda\delta\kappa\tau}
D^{\lambda\delta}_{\quad \alpha\beta} = \alpha_1 \beta_2
F_{\kappa\tau}^{\quad\mu} + {i\alpha_1 \beta_3 \over 2}
\epsilon_{\alpha\beta\kappa\tau} \widetilde F^{\alpha\beta,\mu}\,; \\
&& 2\alpha_2 \beta_4 T_\kappa^{\quad\mu\nu} +i\alpha_3 \beta_7
\epsilon^{\alpha\beta\mu\nu} \widetilde T_{\kappa,\alpha\beta}
=  {\alpha_1 \beta_1 \over m} (\partial^\mu G_\kappa^{\quad \nu}
- \partial^\nu G_\kappa^{\quad\mu})\,; \\
&& 2\alpha_2 \beta_5 R_{\kappa\tau}^{\quad\mu\nu} +i\alpha_3 \beta_8
\epsilon^{\alpha\beta\mu\nu} \widetilde D_{\kappa\tau,\alpha\beta}
+i\alpha_2 \beta_6 \epsilon_{\alpha\beta\kappa\tau} \widetilde
R^{\alpha\beta,\mu\nu}
- {\alpha_3 \beta_9\over 2} \epsilon^{\alpha\beta\mu\nu}
\epsilon_{\lambda\delta\kappa\tau} D^{\lambda\delta}_{\quad \alpha\beta}
= \nonumber\\
&=& {\alpha_1 \beta_2 \over m} (\partial^\mu F_{\kappa\tau}^{\quad \nu}
-\partial^\nu F_{\kappa\tau}^{\quad\mu} ) + {i\alpha_1 \beta_3 \over 2m}
\epsilon_{\alpha\beta\kappa\tau} (\partial^\mu \widetilde
F^{\alpha\beta,\nu} - \partial^\nu \widetilde F^{\alpha\beta,\mu} )\, .
\label{f}
\end{eqnarray}
The essential constraints are:
\begin{eqnarray}
&&\alpha_1 \beta_1 G^\mu_{\quad\mu} = 0\, ,\quad \alpha_1
\beta_1 G_{[\kappa\mu]} = 0 \, ;  
2i\alpha_1 \beta_2 F_{\alpha\mu}^{\quad\mu} +
\alpha_1 \beta_3
\epsilon^{\kappa\tau\mu}_{\quad\alpha} \widetilde F_{\kappa\tau,\mu} =
0\, ;\\
&&2i\alpha_1 \beta_3 \widetilde F_{\alpha\mu}^{\quad\mu}
+ \alpha_1 \beta_2
\epsilon^{\kappa\tau\mu}_{\quad\alpha} F_{\kappa\tau,\mu} = 0\, ;
2i\alpha_2 \beta_4 T^{\mu}_{\quad\mu\alpha} -
 \alpha_3 \beta_{7}
\epsilon^{\kappa\tau\mu}_{\quad\alpha} \widetilde T_{\kappa,\tau\mu}
= 0\, ;\\
&& 2i\alpha_3 \beta_{7} \widetilde
T^{\mu}_{\quad\mu\alpha} -
\alpha_2 \beta_4 \epsilon^{\kappa\tau\mu}_{\quad\alpha}
T_{\kappa,\tau\mu} = 0\, ;\\
&& i\epsilon^{\mu\nu\kappa\tau} \left [ \alpha_2 \beta_6 \widetilde
R_{\kappa\tau,\mu\nu} + \alpha_3 \beta_{8} \widetilde
D_{\kappa\tau,\mu\nu} \right ] + 2\alpha_2 \beta_5
R^{\mu\nu}_{\quad\mu\nu}  + 2\alpha_3
\beta_{9} D^{\mu\nu}_{\quad \mu\nu}  = 0\, ;\\
&& i\epsilon^{\mu\nu\kappa\tau} \left [ \alpha_2 \beta_5 R_{\kappa\tau,
\mu\nu} + \alpha_3 \beta_{9} D_{\kappa\tau, \mu\nu} \right ]
+ 2\alpha_2 \beta_6 \widetilde R^{\mu\nu}_{\quad\mu\nu}
+ 2\alpha_3 \beta_{8} \widetilde D^{\mu\nu}_{\quad\mu\nu}  =0\, ;\\
&& 2i \alpha_2 \beta_5 R_{\beta\mu}^{\quad\mu\alpha} + 2i\alpha_3
\beta_{9} D_{\beta\mu}^{\quad\mu\alpha} + \alpha_2 \beta_6
\epsilon^{\nu\alpha}_{\quad\lambda\beta} \widetilde
R^{\lambda\mu}_{\quad\mu\nu} +\alpha_3 \beta_{8}
\epsilon^{\nu\alpha}_{\quad\lambda\beta} \widetilde
D^{\lambda\mu}_{\quad \mu\nu} = 0\, ;\\
&&2i\alpha_1 \beta_2 F^{\lambda\mu}_{\quad\mu} - 2 i \alpha_2 \beta_4
T_\mu^{\quad\mu\lambda} + \alpha_1 \beta_3 \epsilon^{\kappa\tau\mu\lambda}
\widetilde F_{\kappa\tau,\mu} +\alpha_3 \beta_7
\epsilon^{\kappa\tau\mu\lambda} \widetilde T_{\kappa,\tau\mu} =0\, ;\\
&&2i\alpha_1 \beta_3 \widetilde F^{\lambda\mu}_{\quad\mu} - 2 i \alpha_3
\beta_7 \widetilde T_\mu^{\quad\mu\lambda} + \alpha_1 \beta_2
\epsilon^{\kappa\tau\mu\lambda} F_{\kappa\tau,\mu} +\alpha_2
\beta_4 \epsilon^{\kappa\tau\mu\lambda}  T_{\kappa,\tau\mu} =0\, ;\\
&&\alpha_1 \beta_1 (2G^\lambda_{\quad\alpha} - g^\lambda_{\quad\alpha}
G^\mu_{\quad\mu} ) - 2\alpha_2 \beta_5 (2R^{\lambda\mu}_{\quad\mu\alpha}
+2R_{\alpha\mu}^{\quad\mu\lambda} + g^\lambda_{\quad\alpha}
R^{\mu\nu}_{\quad\mu\nu}) +\nonumber\\
&+& 2\alpha_3 \beta_9
(2D^{\lambda\mu}_{\quad\mu\alpha} + 2D_{\alpha\mu}^{\quad\mu\lambda}
+g^\lambda_{\quad\alpha} D^{\mu\nu}_{\quad\mu\nu})+
2i\alpha_3 \beta_8 (\epsilon_{\kappa\alpha}^{\quad\mu\nu}
\widetilde D^{\kappa\lambda}_{\quad\mu\nu} -
\epsilon^{\kappa\tau\mu\lambda} \widetilde D_{\kappa\tau,\mu\alpha}) -\nonumber\\
&-& 2i\alpha_2 \beta_6 (\epsilon_{\kappa\alpha}^{\quad \mu\nu}
\widetilde R^{\kappa\lambda}_{\quad\mu\nu} -
\epsilon^{\kappa\tau\mu\lambda} \widetilde R_{\kappa\tau,\mu\alpha})
= 0\, ; \\
&& 2\alpha_3 \beta_8 (2\widetilde D^{\lambda\mu}_{\quad\mu\alpha} + 2
\widetilde D_{\alpha\mu}^{\quad\mu\lambda} +g^\lambda_{\quad\alpha}
\widetilde D^{\mu\nu}_{\quad\mu\nu}) - 2\alpha_2 \beta_6 (2\widetilde
R^{\lambda\mu}_{\quad\mu\alpha} +2 \widetilde
R_{\alpha\mu}^{\quad\mu\lambda} + \nonumber\\
&+& g^\lambda_{\quad\alpha} \widetilde
R^{\mu\nu}_{\quad\mu\nu}) +
+ 2i\alpha_3 \beta_9 (\epsilon_{\kappa\alpha}^{\quad\mu\nu}
D^{\kappa\lambda}_{\quad\mu\nu}  - \epsilon^{\kappa\tau\mu\lambda}
D_{\kappa\tau,\mu\alpha} ) -\nonumber\\
&-& 2i\alpha_2 \beta_5
(\epsilon_{\kappa\alpha}^{\quad\mu\nu} R^{\kappa\lambda}_{\quad\mu\nu}
- \epsilon^{\kappa\tau\mu\lambda} R_{\kappa\tau,\mu\alpha} ) =0\, ;\\
&&\alpha_1 \beta_2 (F^{\alpha\beta,\lambda} - 2F^{\beta\lambda,\alpha}
+ F^{\beta\mu}_{\quad\mu}\, g^{\lambda\alpha} - F^{\alpha\mu}_{\quad\mu}
\, g^{\lambda\beta} ) - \nonumber\\
&-&\alpha_2 \beta_4 (T^{\lambda,\alpha\beta}
-2T^{\beta,\lambda\alpha} + T_\mu^{\quad\mu\alpha} g^{\lambda\beta} -
T_\mu^{\quad\mu\beta} g^{\lambda\alpha} ) +\nonumber\\
&+&{i\over 2} \alpha_1 \beta_3 (\epsilon^{\kappa\tau\alpha\beta}
\widetilde F_{\kappa\tau}^{\quad\lambda} +
2\epsilon^{\lambda\kappa\alpha\beta} \widetilde F_{\kappa\mu}^{\quad\mu} +
2 \epsilon^{\mu\kappa\alpha\beta} \widetilde F^\lambda_{\quad\kappa,\mu})
-\nonumber\\
&-& {i\over 2} \alpha_3 \beta_7 ( \epsilon^{\mu\nu\alpha\beta} \widetilde
T^{\lambda}_{\quad\mu\nu} +2 \epsilon^{\nu\lambda\alpha\beta} \widetilde
T^\mu_{\quad\mu\nu} +2 \epsilon^{\mu\kappa\alpha\beta} \widetilde
T_{\kappa,\mu}^{\quad\lambda} ) =0\, .
\end{eqnarray}
They are  the results of contractions of the field function (\ref{ffn1})
with three antisymmetric matrices, as above. Furthermore,
one should recover the relations (\ref{b1}-\ref{f1}) in the particular
case when $\alpha_3 = \beta_3 =\beta_6 = \beta_9 = 0$ and
$\alpha_1 = \alpha_2 = \beta_1 =\beta_2 =\beta_4
=\beta_5 = \beta_7 =\beta_8 =1$.

As a discussion we note that in such a framework we already have physical
content because only certain combinations of field functions
would be equal to zero. In general, the fields
$F_{\kappa\tau}^{\quad\mu}$, $\widetilde F_{\kappa\tau}^{\quad\mu}$,
$T_{\kappa}^{\quad\mu\nu}$, $\widetilde T_{\kappa}^{\quad\mu\nu}$, and
$R_{\kappa\tau}^{\quad\mu\nu}$,  $\widetilde
R_{\kappa\tau}^{\quad\mu\nu}$, $D_{\kappa\tau}^{\quad\mu\nu}$, $\widetilde
D_{\kappa\tau}^{\quad\mu\nu}$ can  correspond to different physical states
and the equations above describe oscillations one state to another.
Furthermore, from the set of equations (\ref{b}-\ref{f}) one
obtains the {\it second}-order equation for symmetric traceless tensor of
the second rank ($\alpha_1 \neq 0$, $\beta_1 \neq 0$):
\begin{equation} {1\over m^2} \left [\partial_\nu
\partial^\mu G_\kappa^{\quad \nu} - \partial_\nu \partial^\nu
G_\kappa^{\quad\mu} \right ] =  G_\kappa^{\quad \mu}\, .
\end{equation}
After the contraction in indices $\kappa$ and $\mu$ this equation is
reduced to the set
\begin{eqnarray}
&&\partial_\mu G_{\quad\kappa}^{\mu} = F_\kappa\,  \\
&&{1\over m^2} \partial_\kappa F^\kappa = 0\, ,
\end{eqnarray}
i.~e.,  to the equations connecting the analogue of the energy-momentum
tensor and the analogue of the 4-vector potential. 
Further investigations may provide additional foundations to
``surprising" similarities of gravitational and electromagnetic
equations in the low-velocity limit, refs.~\cite{Wein2,Jef}.

\noindent
{\bf Acknowledgements.} I am grateful to participants of the recent conferences for useful discussions.

\end{document}